\documentclass[10pt,conference]{IEEEtran} 

\ifCLASSOPTIONcompsoc
  \usepackage[nocompress]{cite}
\else
  \usepackage{cite}
\fi
\ifCLASSINFOpdf
\else
\fi

\usepackage{listings} 
\usepackage{booktabs}
\usepackage{array}
\usepackage{multirow}
\usepackage{url}
\usepackage{graphicx}
\usepackage{amsmath}
\usepackage{tabularx}
\usepackage{makecell}
\usepackage[table]{xcolor}

\definecolor{keyword}{RGB}{255, 0, 0}
\definecolor{string}{RGB}{0, 0, 255}
\definecolor{comment}{RGB}{0, 128, 0}
\definecolor{background}{RGB}{240, 240, 240}

\begin{document}
\title{From Generalist to Specialist: Exploring CWE-Specific Vulnerability Detection}
\author{
    \IEEEauthorblockN{Syafiq Al Atiiq\IEEEauthorrefmark{1}, Christian Gehrmann\IEEEauthorrefmark{2}, Kevin Dahlén\IEEEauthorrefmark{3}, Karim Khalil\IEEEauthorrefmark{4}}
    \IEEEauthorblockA{
        \textit{Lund University}\\
        Lund, Sweden \\
        \{\IEEEauthorrefmark{1}syafiq\_al.atiiq, \IEEEauthorrefmark{2}christian.gehrmann, \IEEEauthorrefmark{4}karim.khalil\}@eit.lth.se, \IEEEauthorrefmark{3}ke8683da-s@student.lu.se 
    }
}

\maketitle
\begin{abstract}
Vulnerability Detection (VD) using machine learning faces a significant challenge: the vast diversity of vulnerability types. Each Common Weakness Enumeration (CWE) represents a unique category of vulnerabilities with distinct characteristics, code semantics, and patterns. Treating all vulnerabilities as a single label with a binary classification approach may oversimplify the problem, as it fails to capture the nuances and context-specific to each CWE. As a result, a single binary classifier might merely rely on superficial text patterns rather than understanding the intricacies of each vulnerability type. Recent reports showed that even the state-of-the-art Large Language Model (LLM) with hundreds of billions of parameters struggles to generalize well to detect vulnerabilities. Our work investigates a different approach that leverages CWE-specific classifiers to address the heterogeneity of vulnerability types. We hypothesize that training separate classifiers for each CWE will enable the models to capture the unique characteristics and code semantics associated with each vulnerability category. To confirm this, we conduct an ablation study by training individual classifiers for each CWE and evaluating their performance independently. Our results demonstrate that CWE-specific classifiers outperform a single binary classifier trained on all vulnerabilities. Building upon this, we explore strategies to combine them into a unified vulnerability detection system using a multiclass approach. Even if the lack of large and high-quality datasets for vulnerability detection is still a major obstacle, our results show that multiclass detection can be a better path toward practical vulnerability detection in the future. All our models and code to produce our results are open-sourced.
\end{abstract}
\IEEEpeerreviewmaketitle
\section{Introduction}
Software vulnerability detection is a resource-intensive task due to the complexity and diversity of software systems \cite{9108283}. Crowdstrike 2024 State of Application Security Report\footnote{\url{https://www.crowdstrike.com/2024-state-of-application-security-report/}} mentioned that \textit{traditional security reviews are time-consuming and expensive}. Vulnerabilities can arise from various sources, such as design flaws, implementation errors, or misconfigurations. Traditional vulnerability detection techniques, such as static analysis \cite{1366126} and dynamic testing \cite{ball1999concept}, often struggle to keep pace with the rapidly evolving threat landscape and the increasing complexity of software. These techniques can be time-consuming, resource-intensive, and may miss certain types of vulnerabilities. Moreover, the expertise required to identify and mitigate vulnerabilities effectively is often in short supply \cite{burrell2020exploration}, making it difficult for organizations to maintain a robust security posture in a timely manner.

Recent advancements in transformers \cite{NIPS2017_3f5ee243}, particularly in large language models (LLMs), have shown remarkable capabilities in understanding and generating human-like code. This has led to the consideration of leveraging LLMs for vulnerability analysis. Linus Torvalds, the creator of the Linux Kernel, anticipates that LLMs can help developers catch bugs more easily, as many of the bugs he encountered during the development did not require the expertise of a senior developer\footnote{\url{https://www.youtube.com/watch?v=w7-gJicosyA}}. This suggests that LLMs could be suitable for addressing low-hanging fruit problems in software development. Moreover, LLMs have demonstrated impressive abilities in comprehending and generating both natural language \cite{NEURIPS2020_1457c0d6} and programming languages \cite{wei2023magicoder,guo2024deepseekcoder,zheng2024opencodeinterpreter}. By training LLMs on vast amounts of code and their associated security-related metadata, they can potentially learn patterns, best practices, and common vulnerabilities. This makes LLMs a promising tool for automated code analysis and vulnerability detection. Consequently, the question arises as to whether LLMs can be effectively utilized to identify and mitigate software vulnerabilities, potentially reducing the reliance on manual efforts and improving the efficiency of vulnerability detection processes.

Diversevul \cite{Diversevul} has shown the promising potential of using LLMs for vulnerability detection by finetuning smaller language models such as RoBERTa \cite{liu2019roberta} based models (CodeBERT \cite{feng2020codebert} and GraphCodeBERT \cite{guo2021graphcodebert}), GPT-2 \cite{radford2019language} based models (CodeGPT \cite{lu2021codexglue} and PolyCoder \cite{10.1145/3520312.3534862}), and T5 \cite{10.5555/3455716.3455856} based models (CodeT5 \cite{wang-etal-2021-codet5} and NatGen \cite{10.1145/3540250.3549162}). These models were trained on a curated dataset of 330,492 non-vulnerable and 18,945 vulnerable functions, resulting in significant improvements over traditional graph neural network (GNN) \cite{4700287} approaches. However, while LLMs outperform GNN-based classifiers, they struggle to generalize well when predicting vulnerabilities in unseen projects, resulting in extremely low F1 scores. Diversevul's best-finetuned model (based on CodeBERT \cite{feng2020codebert}), trained on their own data combined with all previously available data, achieved only an 11.94\% F1 score at best, with the cause for this issue being unclear.

As a follow-up, PrimeVul\cite{ding2024vulnerabilitydetectioncodelanguage} is introduced as a new benchmark dataset designed to show the limitation of the existing vulnerability detection problem. PrimeVul features high-quality labeled data, a rigorous de-duplication process, and realistic evaluation settings, setting it apart from previous datasets. The PrimeVul study reveals a deeper problem of VD: even state-of-the-art LLMs with billions of parameters struggle to generalize well, with performance significantly lower than reported on previous benchmarks. For instance, a StarCoder2 \cite{lozhkov2024starcoder2stackv2} with a 7B parameter model achieved an F1 score of 68.26\% on the widely-used BigVul \cite{bigvul} dataset but only 3.09\% on PrimeVul. Furthermore, the PrimeVul study introduces new evaluation guidelines, such as the Vulnerability Detection Score (VD-S) and pair-wise evaluation, which provide a more realistic assessment of model performance in real-world scenarios. 

In this paper, we investigate a different way to improve the practical applicability of LLMs for code vulnerability detection in a real-world setting. During our analysis, we observed two key issues. First, the diversity of characteristics, code semantics, and patterns between CWE classifications is too diverse to treat as a single label. For example, CWE-89\footnote{\url{https://cwe.mitre.org/data/definitions/89.html}}: Improper Neutralization of Special Elements used in an SQL Command (``SQL Injection'') involves user-supplied input being passed unsanitized into SQL queries, allowing an attacker to manipulate the query's structure and execute unintended commands. This vulnerability arises from the lack of proper input validation and sanitization. In contrast, another CWE category, such as CWE-125\footnote{\url{https://cwe.mitre.org/data/definitions/125.html}}: Out-of-bounds Read, involves accessing memory outside the bounds of an allocated buffer, which stems from improper bounds checking. Treating these diverse vulnerabilities, each with its own unique code semantics and patterns, as a single ``vulnerable code'' label might oversimplify the problem and prevent models from learning the specific nuances of each vulnerability type. 

Second, the CWE representation data suffers from a severe imbalance problem. Some CWE labels are significantly more frequent than others, with the most common label accounting for a substantial portion of the dataset. In contrast, the least common label has only a few instances. Given this imbalance, even if a model is able to generalize over multiple CWEs, it would likely learn some CWEs better than others. The model might become biased towards the most frequent CWE types, as they dominate the training data while struggling to effectively learn and detect the less common ones.

To analyze these issues even further, we conduct an ablation study with two following research questions:

\begin{itemize}
\item RQ1: How do CWE-specific classifiers perform compared to a single binary classifier in detecting vulnerabilities? (The definition of a CWE-specific classifier is a vulnerability detector that is trained on a specific CWE category). If the CWE-specific classifier performs better:
\item RQ2: Can treating vulnerability detection as a multi-class problem, with each CWE as a separate class, improve detection performance?
\end{itemize}

The rest of the paper is organized as follows. We present the background and related work in section \ref{related_work}. We elaborate on the problem of the current vulnerability detection system in section \ref{problem}. We explain the data collection and processing in section \ref{data}. We perform our experiment, evaluation, and discussion in \ref{experimental}. And finally, we draw our conclusion and anticipate future works in section \ref{conclusion}.

\section{Background \& Related Work}
\label{related_work}
This section provides an overview of the background and related work on vulnerability detection using machine learning techniques. We begin by introducing the CWE, a widely used taxonomy for categorizing software vulnerabilities, and discuss its importance. We then review the evolution of machine learning-based approaches for vulnerability detection, from early work using traditional machine learning algorithms to more recent advancements leveraging deep learning and pre-trained language models. Finally, we survey several datasets for training and evaluating machine learning models for vulnerability detection and highlight recent findings on data quality issues in these datasets.

\subsection{Common Weakness Enumeration (CWE)}
\label{cwe}
CWE is a community-developed list of common software and hardware weakness types that have security ramifications \cite{christey2009cwe}. CWE serves as a common language, a measuring stick for security tools, and a baseline for weakness identification, mitigation, and prevention efforts \cite{10.1145/1387830.1387835}. CWE provides a standardized, unified vocabulary for discussing and describing software weaknesses. It enables more effective communication, knowledge sharing, and collaboration among software developers, security professionals, and organizations. CWE also helps in assessing and prioritizing risks associated with different types of vulnerabilities, guiding secure coding practices, and improving the overall security of software systems.

CWE classifies software weaknesses based on several factors, including:

\begin{itemize}
\item \textbf{Abstraction}: CWE categorizes weaknesses at different levels of abstraction, from high-level architectural flaws to low-level coding errors.
\item \textbf{Scope}: Weaknesses are classified based on their scope of impact, i.e., system, language, or technology specific.
\item \textbf{Consequences}: CWE considers the potential consequences of weaknesses, such as information disclosure, denial of service, or code execution.
\item \textbf{Domains}: Weaknesses are categorized based on the domains they affect, such as authentication, cryptography, or input validation.
\end{itemize}

Each weakness in the CWE list is assigned a unique identifier, known as a CWE ID. It is a number that follows the format ``CWE-\textit{X}'', where \textit{X} is a sequential number.

\subsection{Vulnerability Detection with Machine Learning}

Static code analysis has long been a method for identifying vulnerabilities and ensuring code quality. Traditional static analysis techniques rely on rule-based systems and heuristics to detect common programming errors and security flaws. However, with the advent of machine learning, researchers have explored the application of these techniques to enhance the effectiveness and efficiency of static code analysis.

Early work \cite{10.1145/1315245.1315311} proposed using support vector machines (SVMs) \cite{cortes1995support} to predict vulnerable software components by extracting various features from code to identify components containing vulnerabilities based on historical data. VulDeePecker \cite{LiZXO0WDZ18} advanced this approach by applying deep learning techniques, specifically long short-term memory (LSTM) \cite{6795963} networks, to learn the patterns and representations of secure and insecure code. They treated code as a sequence of tokens and trained the LSTM model to predict the presence of vulnerabilities based on the learned code representations.

Building upon these, more recent work has focused on leveraging Graph Neural Networks (GNNs) \cite{4700287} to incorporate code structure information into vulnerability detection. Devign \cite{devign} and ReVeal \cite{reveal} utilized GNNs with code property graphs, while VulChecker \cite{285507} proposed an enriched program dependence graph. Similarly, \cite{6956589} proposed a GNN-based approach to model code as a graph, capturing the relationships between code elements and identifying vulnerabilities based on the structural and semantic properties of the code.

Prior to the widespread use of pre-trained language models, early deep learning approaches for vulnerability detection primarily utilized Bidirectional Long Short-Term Memory (BiLSTM) and Bidirectional Gated Recurrent Unit (BiGRU) \cite{cho-etal-2014-learning} networks. These variants of recurrent neural networks (RNNs) can process sequential data, such as code tokens, in both forward and backward directions, enabling them to learn the patterns and representations of code for vulnerability detection tasks. Notable examples include VulDeePecker \cite{LiZXO0WDZ18}, a BiLSTM-based system for vulnerability detection, and SySeVR \cite{9321538}, a framework that utilizes both BiLSTM and BiGRU models to detect software vulnerabilities.

Recent advancements in natural language processing have led to the adoption of pre-trained language models, such as Bidirectional Encoder Representations from Transformers (BERT) \cite{devlin2019bert}, in the field of code analysis. CodeBERT \cite{feng2020codebert} is a pre-trained model for programming languages that adapts the BERT architecture to capture the semantic and syntactic properties of code. While CodeBERT is not specifically designed for vulnerability detection, its ability to understand code semantics suggests its potential application in this domain.

\cite{10.1145/3564625.3567985} compared the performance of transformer-based language models, including BERT, GPT-2, and their variants, against BiLSTM and BiGRU models on a dataset spanning CWE-119 and CWE-399 in C/C++ code. They found that the transformer models outperformed the RNN-based models in detecting vulnerabilities, achieving higher precision, recall, and F1 scores. They attribute this to the transformer models' ability to capture long-range dependencies and learn more expressive code representations through self-attention mechanisms. The former refers to the ability of a model to capture and understand relationships between elements that are far apart in a sequence. The latter enables the model to weigh the importance of different elements in the input sequence based on their relevance to each other, regardless of their distance.

Similar to \cite{10.1145/3564625.3567985}, Diversevul \cite{Diversevul} conducted a comprehensive study on 11 model architectures from 4 families: GNNs \cite{4700287}, RoBERTa \cite{liu2019roberta}, GPT-2 \cite{radford2019language}, and T5 \cite{10.5555/3455716.3455856}. Their results demonstrated that LLMs, especially those pre-trained with code-specific tasks like CodeT5 \cite{wang2021codet5,le2022coderl,wang2023codet5plus} and NatGen \cite{10.1145/3540250.3549162}, significantly outperformed the state-of-the-art GNN model as the training data size increased.

Diversevul \cite{Diversevul} has confirmed that on extensive datasets, even LLMs with code-specific capabilities suffer from a low F1 score, indicating they cannot generalize well to unseen code. It mentioned that the cause is unclear, with one possible reason for model overfitting. As a follow-up, PrimeVul \cite{ding2024vulnerabilitydetectioncodelanguage}, a new benchmark dataset designed to address the limitations of existing vulnerability datasets, is introduced. It employs rigorous data de-duplication and chronological data splitting strategies to mitigate data leakage. Furthermore, it introduces more realistic evaluation metrics, such as the Vulnerability Detection Score (VD-S), and a pair-wise evaluation method to assess the model's ability to distinguish between vulnerable code and its fixed counterpart. By evaluating code language models on PrimeVul, the authors demonstrate that existing benchmarks significantly overestimate the performance of these models in real-world vulnerability detection scenarios. PrimeVul concludes that despite attempts to enhance the performance of code language models through advanced training techniques and larger model architectures, these models still fall significantly short of the requirements for reliable and effective vulnerability detection in real-world scenarios.

\subsection{Vulnerable Code Datasets}
\label{dataset}
Several datasets have been curated in recent years to enable research on using machine learning for automated software vulnerability detection. We survey the most relevant below.

Devign \cite{devign} is a manually labeled dataset of vulnerable and non-vulnerable functions extracted from commits in 4 large real-world C projects. It addresses some shortcomings of synthetic datasets used in prior work but is limited in size due to the manual labeling effort required ($\pm$600 man-hours).

ReVeal \cite{reveal} is a dataset collected from the Chromium and Debian projects. It labels functions changed in security patches as vulnerable and unchanged functions as non-vulnerable. The authors use this dataset to highlight challenges in existing deep learning approaches, i.e., duplication, class imbalance, learning irrelevant features, and inadequate models. They propose a roadmap to address these issues, including using representation learning, data deduplication, and class rebalancing techniques.

BigVul \cite{bigvul}, CrossVul \cite{crossvul}, and CVEfixes \cite{cvefixes} are datasets automatically curated by identifying vulnerability fixing commits from security issue trackers and the Common Vulnerabilities and Exposures (CVE) database. These datasets cover a wide range of open-source projects and provide samples of real-world vulnerable code paired with their fixed versions. CVEfixes is the largest among these, with 8,932 vulnerable functions across 168,089 functions from 564 projects. However, the data labeling approach of considering all functions modified in a vulnerability fixing commit as vulnerable introduces some noise. Building on CVEfixes, DiverseVul \cite{Diversevul} further expands the dataset to 18,945 vulnerable functions spanning 150 CWEs and 330,492 non-vulnerable functions from 797 projects. It is currently the largest and most diverse dataset in this space. The authors also provide an in-depth study benchmarking 11 deep learning model architectures on this dataset and uncover insights about current challenges.

A recent study \cite{10.1109/ICSE48619.2023.00022} investigated the issue of data quality in software vulnerability datasets. They found substantial labeling errors in both the BigVul \cite{bigvul} and Devign \cite{devign} datasets, which is also mentioned in the Diversevul \cite{Diversevul} paper. Specifically, their manual analysis revealed that 45.7\% of BigVul labels and 20\% of Devign labels were inaccurate. Furthermore, both datasets exhibited label inconsistency issues caused by latent vulnerabilities (BigVul) and simultaneous code branches (Devign). The study also found a high prevalence of code duplication in BigVul (17\%) and Devign (10.1\%), which can lead to inflated performance metrics due to data leakage. These findings suggest that the reliability of BigVul and Devign for training and evaluating vulnerability prediction models is questionable, as the noise and potential bias introduced by these data quality issues could lead to untrustworthy model performance. Therefore, to maintain the integrity of our models and results, we decided to omit BigVul and Devign and focus on the four remaining datasets with higher quality.

\subsection{Evaluation Metrics for Vulnerability Detection}
Traditionally, accuracy and F1 scores have been widely used to evaluate the performance of vulnerability detection models. However, these metrics fail to capture the practical utility of such models in real-world scenarios, where the trade-off between false positives and false negatives is crucial [1].

To address this limitation, PrimeVul \cite{ding2024vulnerabilitydetectioncodelanguage} proposes a new evaluation metric called Vulnerability Detection Score (VD-S). VD-S measures the false negative rate (FNR) of a vulnerability detector while ensuring that the false positive rate (FPR) is below a fixed threshold $r$, i.e., $FNR @ (FPR \leq r)$. This metric emphasizes the importance of minimizing false negatives (missed vulnerabilities) while keeping false positives (false alarms) under control, reflecting the practical requirements of deploying vulnerability detection tools. The configurable parameter $r$ in VD-S allows for adjusting the maximum tolerable false positive rate based on the specific application scenario. A lower VD-S indicates better vulnerability detection performance at the given false positive rate tolerance.

\section{The Problem of VD System}
\label{problem}
\subsection{The Diversity of Vulnerable Code}
One of the key challenges in vulnerability detection using code language models is the vast diversity of characteristics, code semantics, and patterns among different CWEs. Treating these distinct characteristics as a single label of ``vulnerable code" might oversimplify the problem and hinder the model's ability to learn the specific nuances of each vulnerability type. To illustrate this point, let's consider two contrasting CWE categories: CWE 89 and 125. It's important to note that these are just two examples among the numerous CWEs that exist.

CWE-89 involves user-supplied input being passed unsanitized into SQL queries, allowing an attacker to manipulate the query's structure and execute unintended commands. This vulnerability arises from the lack of proper input validation and sanitization. Here's an example of CWE-89:

\begin{lstlisting}[language=c]
void vulnerable_function(char *username, char *password) {
    MYSQL *conn;
    char query[200];
    // ... (establish database connection)

    sprintf(query, "SELECT * FROM users WHERE username='%s' AND password='%s'", username, password);

    if (mysql_query(conn, query)) {
        // ... (handle error)
    }
    // ... (process query results)
\end{lstlisting}

In this example, the \texttt{sprintf} function is used to construct the SQL query by directly concatenating the user-supplied \texttt{username} and \texttt{password} into the query string. An attacker can input malicious SQL code as the \texttt{username} or \texttt{password}, such as \texttt{admin' --}, which would alter the query's structure and bypass the password check, potentially granting unauthorized access. To prevent this, it is essential to use parameterized queries or prepared statements instead of constructing queries through concatenation. Additionally, proper input validation and sanitization should be implemented to ensure that user-supplied input is handled securely.

On the other hand, CWE-125 involves accessing memory outside the bounds of an allocated buffer, which stems from improper bounds checking. Here's an example:

\begin{lstlisting}[language=c]
void vulnerable_function(char *user_input) {
    char buffer[10];
    strcpy(buffer, user_input);
    // ...
}
\end{lstlisting}

In this code snippet, the \texttt{strcpy} function is used to copy the contents of \texttt{user\_input} into the fixed-size \texttt{buffer} array. If the length of \texttt{user\_input} exceeds the size of \texttt{buffer}, it will result in a buffer overflow, allowing an attacker to overwrite adjacent memory locations and potentially execute arbitrary code. These examples demonstrate the fundamental differences in characteristics, code semantics, and patterns between just two CWE categories. In reality, there are hundreds of CWE classifications, each with its own unique set of vulnerabilities and code patterns. Treating these diverse vulnerabilities as a single label might lead to several problems:
\begin{itemize}
    \item Oversimplification: By grouping different CWEs under a single label, the model may fail to capture the specific nuances and patterns associated with each category.
    \item Reduced learning effectiveness: The model may struggle to learn the distinct code semantics and characteristics of each vulnerability type, as the training data would lack the necessary granularity and specificity.
    \item Limited interpretability: When the model predicts a code snippet as vulnerable, it would be challenging to determine which specific vulnerability type the model has identified, making it difficult to interpret the results and take appropriate remediation actions.
\end{itemize}

To address these challenges, it is crucial to consider the diversity of CWE classifications when training code language models for vulnerability detection.

\subsection{Imbalance Between Vulnerable and Non-Vulnerable}

Another challenge in vulnerability detection is the imbalance between vulnerable and non-vulnerable code samples. As shown in Table \ref{tab_func}, the number of non-vulnerable code samples (315,478) significantly exceeds the number of vulnerable code samples (25,024) in the entire dataset of 340,502 samples. This imbalance, where non-vulnerable samples account for approximately 92.65\% of the total data, can lead to biased models that struggle to effectively learn and detect vulnerabilities, as the non-vulnerable samples are more dominant.

\subsection{Data Imbalance within the Vulnerable Code}
\begin{figure}
    \centering
    \includegraphics[width=1\linewidth]{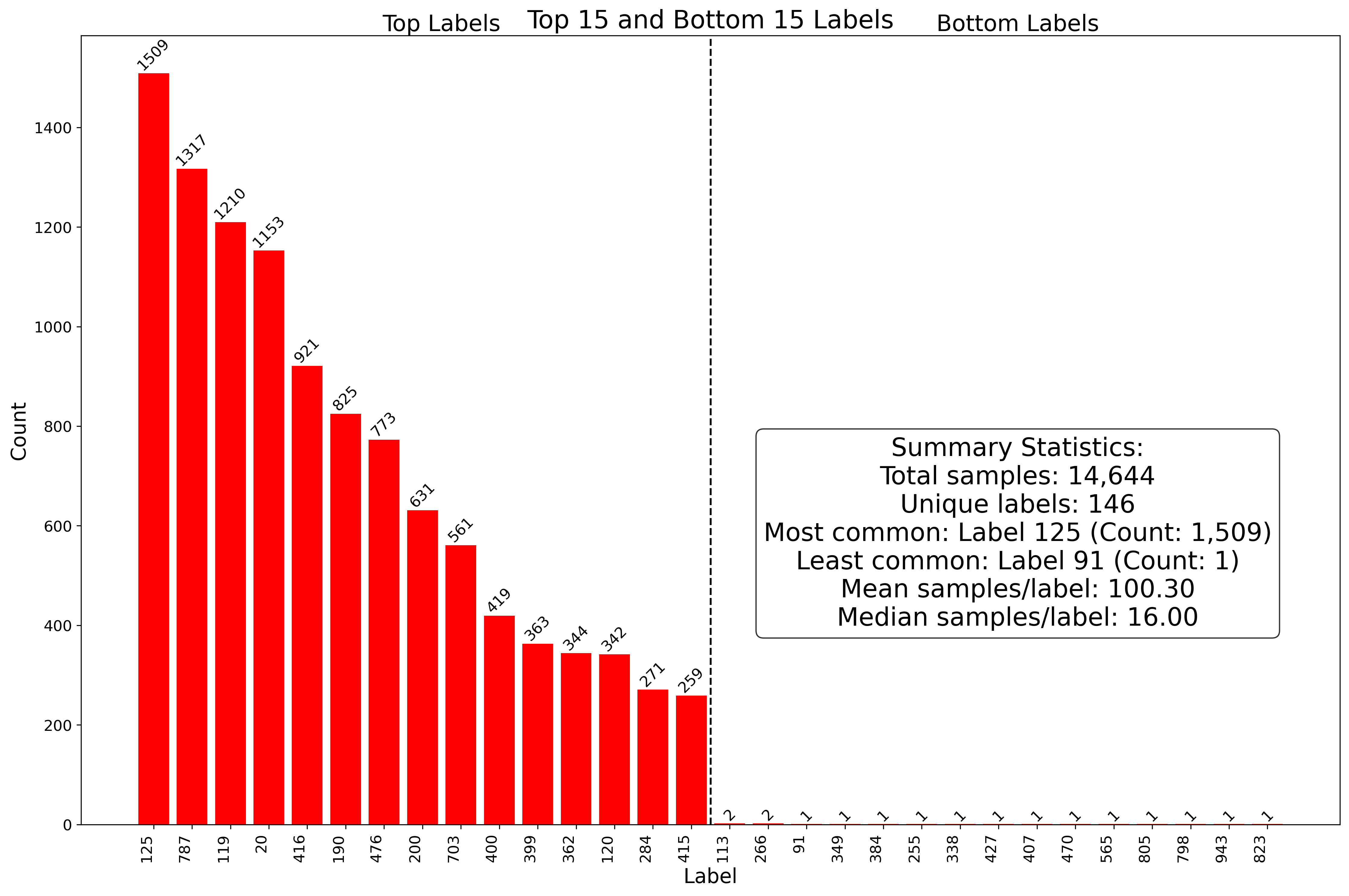}
    \caption{Top and Bottom 15 of CWE IDs in the vulnerable function}
    \label{fig_cwe_dist}
\end{figure}

In addition to the imbalance between vulnerable and non-vulnerable code, there is also a significant imbalance within the vulnerable code samples across different CWE categories. Figure \ref{fig_cwe_dist} illustrates the distribution of the top and bottom 15 CWE categories in the vulnerable code samples. It is clear that some CWEs, such as 119 and 79, have a significantly higher number of samples compared to others, such as 805 and 134.

This imbalance within the vulnerable code can lead to models that are biased towards the most frequent CWE categories, as they have more representative samples in the training data. Consequently, the models may struggle to effectively learn and detect vulnerabilities in the less common CWE categories.

\section{Data Collection \& Processing}
\label{data} 
In this section, we detail the data processing steps applied to the datasets used in studying our research questions. We also present an analysis of the preprocessed data to provide insights into its characteristics and distribution. The code used for data preprocessing and analysis is available in our repository\footnote{\url{https://anonymous.4open.science/r/cwe_vd-7BC6/}}. Not all the datasets contain information regarding the CWE classification number. Among the previously mentioned datasets, only the following datasets contain the CWE number as their features: Diversevul \cite{Diversevul}, CrossVul \cite{crossvul}, and CVEFixes \cite{cvefixes}. ReVeal \cite{reveal} does not meet the requirement as it only contains the label for vulnerable or not.

The DiverseVul\cite{Diversevul} dataset is a significant component of our study due to its comprehensive coverage of a wide range of vulnerability types. However, unlike the approach used in PrimeVul\cite{ding2024vulnerabilitydetectioncodelanguage}, where the train:test split is based on the year (i.e., earlier years for training and later years for testing), we do not employ this method due to the lack of complete year information in DiverseVul. Despite not splitting the data based on years, we expect our results to be roughly the same as those obtained using a year-based split. This is because our dataset has been thoroughly de-duplicated, ensuring that there is no data leakage between the training and testing datasets. The de-duplication process helps maintain the integrity of the evaluation, as it prevents the model from learning from examples that are present in both the training and testing sets.

Note that CVEFixes provides a script to scrape the data. Since the original data is only until 9 June 2021\footnote{https://github.com/secureIT-project/CVEfixes}, we scrape the remaining data until the day this paper is written. Similar to PrimeVul \cite{ding2024vulnerabilitydetectioncodelanguage}, we employ a multi-step process to eliminate duplicate functions from the dataset. Initially, we normalize the functions by each commit by removing whitespace characters such as spaces, tabs (``$\backslash t$"), newlines (``$\backslash n$"), and carriage returns (``$\backslash r$"). Subsequently, we calculate the MD5 hash for each function. All the functions are then aggregated, and a de-duplication is performed using the MD5 hashes of the normalized functions. During this process, a set of unique hashes is maintained, and if the hash of a normalized function already exists in the set, that function is excluded from further processing. This approach ensures that the resulting dataset contains only unique and changed functions, effectively reducing redundancy and noise in the data.

\begin{figure}
    \centering
    \includegraphics[width=1\linewidth]{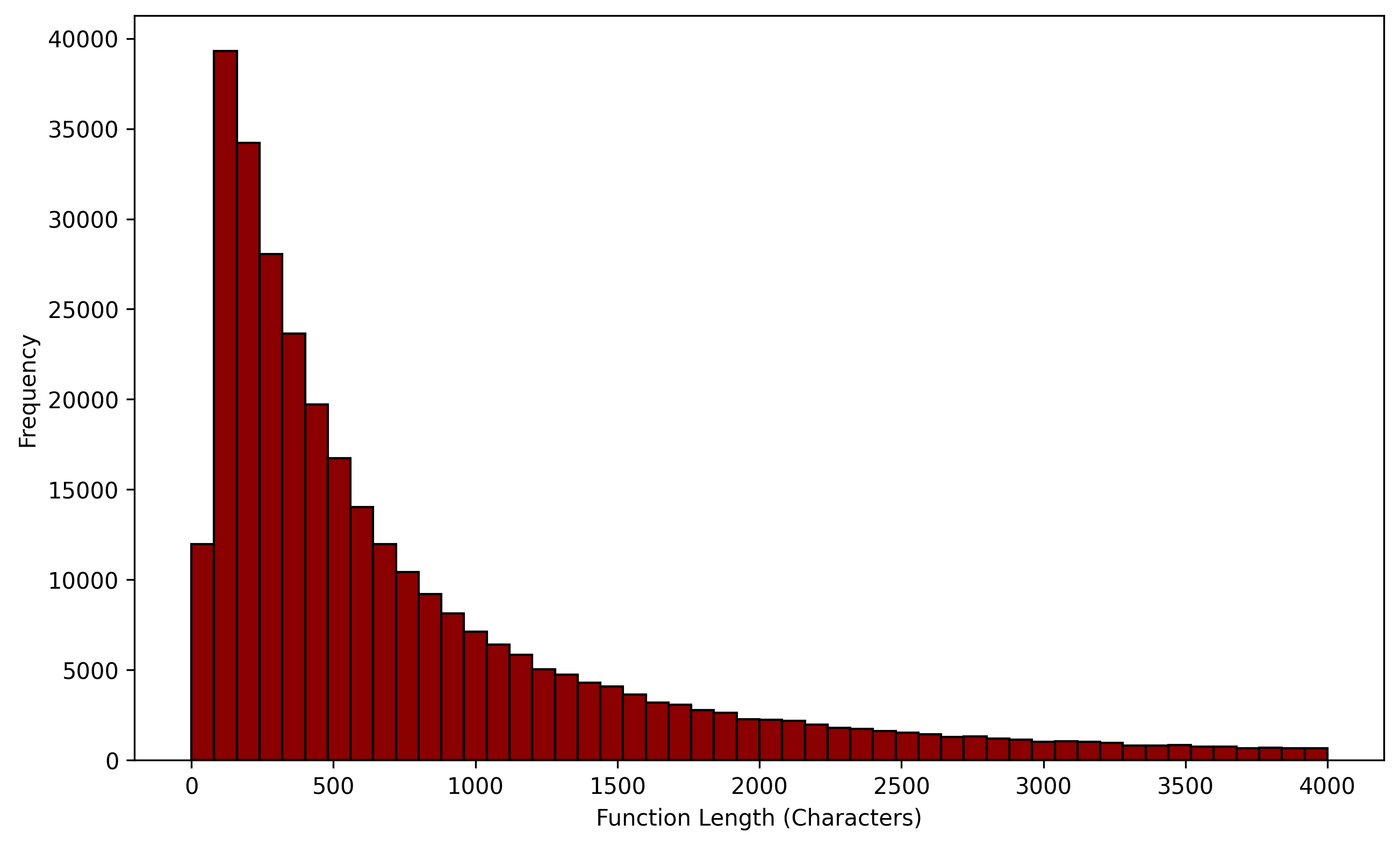}
    \caption{Final data distribution, after filtered (account for 91.79\% of the total data)}
    \label{fig_func_dist_filtered}
\end{figure}

Taking into account the computational benefits and the distribution of the merged dataset, we decided to set a maximum function length of 4,000 characters and exclude the remaining data that exceeds this limit. This decision was based on the observation that the majority of the data (91.9\%) falls within this 4,000-character threshold, as shown in Table \ref{tab_func}. Figure \ref{fig_func_dist_filtered} presents the distribution of function lengths in the filtered dataset, revealing a highly skewed, long-tailed distribution. The majority of functions have relatively short lengths, with a rapid decrease in the number of functions as the length increases. Beyond the 1,500-character point, the occurrence of functions becomes more sparse and scattered, indicating that longer functions are less common in the dataset and may represent more complex or specialized code snippets. 

Finally, the source dataset can be expressed as:

\begin{equation}
\begin{aligned}
    d = \{x \mid x \in & (\text{DV} \cup \text{CF} \cup \text{CV}), \text{len}(x) \leq 4000, \text{count}(x) = 1\}
\end{aligned}
\end{equation}

Where DV = DiverseVul, CF = CVEFixes, and CV = CrossVul.

\begin{table}[ht]
    \centering
    \caption{Dataset's length distribution}
    \label{tab_func}
    \begin{tabular}{@{}lccccc@{}}
        \toprule
        & \multicolumn{4}{c}{\textbf{Code length}} \\
        \cmidrule(lr){2-5}
        \textbf{Label}       & \textbf{$l<$4k}    & \textbf{4k$\leq l<$8k} & \textbf{8k$\leq l<$12k} & \textbf{$l\geq$12k} & \textbf{All}    \\
        \midrule
        Vuln          & \cellcolor{yellow}16,955 & 2,768 & 1,071 & 4,230 & 25,024 \\
        Non-Vuln      & \cellcolor{yellow}295,587 & 11,242 & 2,802 & 5,847 & 315,478 \\
        \textbf{Total}      & \cellcolor{yellow}\textbf{312,542} & \textbf{14,010} & \textbf{3,873} & \textbf{10,077} & \textbf{340,502} \\
        \textbf{\% Total}   & \cellcolor{yellow}\textbf{91.79} & \textbf{4.11} & \textbf{1.14} & \textbf{2.96} & \textbf{100} \\
        \bottomrule
    \end{tabular}
\end{table}

The dataset $d$ used in this study consists of 312,542 code samples, as highlighted in Table \ref{tab_func}. It is composed of two subsets: $d_v$, which contains vulnerable code samples, and $d_{nv}$, which contains non-vulnerable code samples. We split the dataset based on the CWE number associated with each vulnerable code sample in $d_v$. This approach guarantees that the distribution of CWE numbers in the training and testing sets is consistent with the overall dataset.

We employ a 90:10 split ratio for training and testing, respectively. Let $D_{CWE-x}$ denote the set of vulnerable code samples in $d_v$ associated with a specific CWE number $x$. For each unique CWE number in $d_v$, we randomly select 90\% of the code samples from $D_{CWE-x}$ to form the vulnerable training set $d_{v-train}$ and the remaining 10\% to form the vulnerable testing set $d_{v-test}$. This process is repeated for all unique CWE numbers in $d_v$. Code samples in $d_v$ that are missing CWE information are neglected.

The non-vulnerable code samples in $d_{nv}$ are also split using the same 90:10 ratio to form the non-vulnerable training set $d_{nv-train}$ and non-vulnerable testing set $d_{nv-test}$.

Formally, for a dataset $d$ with $n$ unique CWE numbers in $d_v$, the training sets $d_{v-train}$, $d_{nv-train}$ and testing sets $d_{v-test}$, $d_{nv-test}$ are constructed as follows:

$d_{v-train} = \bigcup_{i=1}^{n} {x \in D_{CWE-i} \mid x \in (D_{CWE-i} \xrightarrow{0.9})}$

$d_{v-test} = \bigcup_{i=1}^{n} {x \in D_{CWE-i} \mid x \in (D_{CWE-i} \xrightarrow{0.1})}$

$d_{nv-train} = d_{nv} \xrightarrow{0.9}$

$d_{nv-test} = d_{nv} \xrightarrow{0.1}$

Where $D_{CWE-i} \xrightarrow{p}$ represents the random split operation that selects a subset of code samples from $D_{CWE-i}$ with a proportion of $p$. The overall training set $d_{train}$ is the union of the vulnerable training set $d_{v-train}$ and the non-vulnerable training set $d_{nv-train}$, i.e., $d_{train} = d_{v-train} \cup d_{nv-train}$. Similarly, the overall testing set $d_{test}$ is the union of the vulnerable testing set $d_{v-test}$ and the non-vulnerable testing set $d_{nv-test}$, i.e., $d_{test} = d_{v-test} \cup d_{nv-test}$.

\section{Experimental Results \& Discussions}
\label{experimental}

\subsection{Fine-tuning Setup}
For our experiments, we utilized the DeepSeek-Coder-1.3B-Instruct \cite{guo2024deepseekcoder} model as our base model. This choice was motivated by several factors. The model's relatively small size (1.3B parameters) allowed for efficient fine-tuning within our computational constraints. Despite its compact size, DeepSeek-Coder-1.3B-Instruct demonstrates competitive performance in code understanding and generation tasks, often surpassing models with larger parameters. Readers can refer to the EvalPlus\footnote{https://evalplus.github.io/leaderboard.html} leaderboard for detailed comparisons.

Our fine-tuning process employed a learning rate of 2e-5 and ran for ten epochs. We utilized NVIDIA A100 (80GB) GPU clusters to perform the fine-tuning. The code for our fine-tuning process, including data preprocessing, model training, and evaluation scripts, is available in our repository. We encourage readers interested in reproducing our results or building upon our work to access and utilize these resources.

\subsection{RQ1: How do CWE-specific classifiers perform compared to a single binary classifier in detecting vulnerabilities?}
\begin{table*}[t]
\centering
\begin{tabular}{|l|l|c|c|c|c|c|c|c|c|c|c|}
\hline
Model & Test & VD-S $\downarrow$ & Acc $\uparrow$ & F1 $\uparrow$ & Prec $\uparrow$ & Rec $\uparrow$ & FPR $\downarrow$ & TP $\uparrow$ & TN $\uparrow$ & FP $\downarrow$ & FN $\downarrow$ \\
\hline
$m_{all}$ & $d_{test_{balanced}}$ & 1.0000 & 0.7577 & 0.7789 & 0.7163 & 0.8535 & 0.3380 & 1270 & 985 & 503 & 218 \\
\hline
$m_{125}$ & \multirow{2}{*}{$d_{test_{125}}$} & 0.0867 & 0.8833 & 0.8867 & 0.8616 & 0.9133 & 0.1467 & 137 & 128 & 22 & 13 \\
\cline{1-1}\cline{3-12}
$m_{all}$ &  & 1.0000 & 0.8033 & 0.8218 & 0.7514 & 0.9067 & 0.3000 & 136 & 105 & 45 & 14 \\
\cline{1-1}\cline{2-12}
$m_{787}$ & \multirow{2}{*}{$d_{test_{787}}$} & 0.0909 & 0.8636 & 0.8696 & 0.8333 & 0.9091 & 0.1818 & 120 & 108 & 24 & 12 \\
\cline{1-1}\cline{3-12}
$m_{all}$ &  & 1.0000 & 0.8182 & 0.8298 & 0.7800 & 0.8864 & 0.2500 & 117 & 99 & 33 & 15 \\
\cline{1-1}\cline{2-12}
$m_{119}$ & \multirow{2}{*}{$d_{test_{119}}$} & 0.1417 & 0.8458 & 0.8477 & 0.8374 & 0.8583 & 0.1667 & 103 & 100 & 20 & 17 \\
\cline{1-1}\cline{3-12}
$m_{all}$ &  & 1.0000 & 0.7375 & 0.7586 & 0.7021 & 0.8250 & 0.3500 & 99 & 78 & 42 & 21 \\
\cline{1-1}\cline{2-12}
$m_{20}$ & \multirow{2}{*}{$d_{test_{20}}$} & 0.2368 & 0.7895 & 0.7838 & 0.8056 & 0.7632 & 0.1842 & 87 & 93 & 21 & 27 \\
\cline{1-1}\cline{3-12}
$m_{all}$ &  & 1.0000 & 0.7675 & 0.7888 & 0.7226 & 0.8684 & 0.3333 & 99 & 76 & 38 & 15 \\
\cline{1-1}\cline{2-12}
$m_{416}$ & \multirow{2}{*}{$d_{test_{416}}$} & 0.1630 & 0.8478 & 0.8462 & 0.8556 & 0.8370 & 0.1413 & 77 & 79 & 13 & 15 \\
\cline{1-1}\cline{3-12}
$m_{all}$ &  & 1.0000 & 0.7228 & 0.7385 & 0.6990 & 0.7826 & 0.3370 & 72 & 61 & 31 & 20 \\
\hline
\end{tabular}
\caption{Comparison of CWE-specific classifiers and a single binary classifier for the top 5 CWE categories at an FPR tolerance of $r=0.2$.}
\label{tab_rq1}
\end{table*}

To address RQ1, we conducted experiments comparing the performance of CWE-specific classifiers ($m_{CWE}$) against a single binary classifier ($m_{all}$). We selected the top 5 most common CWEs, as shown in Figure \ref{fig_cwe_dist}. They are CWE: 125, 787, 119, 20, and 416.

For each CWE category, we construct the CWE-specific training set ($d_{train_{CWE}}$) by combining the vulnerable samples from the corresponding CWE category ($D_{CWE}$) taken from $d_{v-train}$ and an equal number of non-vulnerable samples. Similarly, the CWE-specific testing set ($d_{test_{CWE}}$) is constructed by combining the vulnerable samples from the corresponding CWE category ($D_{CWE}$) taken from $d_{v-test}$ and an equal number of non-vulnerable samples. It is important to note that the non-vulnerable samples used in each CWE-specific dataset are disjoint. We train a separate classifier ($m_{CWE}$) using each CWE-specific training set ($d_{train_{CWE}}$) and evaluate its performance on the respective CWE-specific testing set ($d_{test_{CWE}}$).

For the single binary classifier ($m_{all}$), we construct a balanced training set by combining all the vulnerable samples from $d_{v-train}$ and an equal number of non-vulnerable samples randomly selected from $d_{nv-train}$. The balanced testing set for $m_{all}$, denoted as $d_{test_{balanced}}$, is constructed similarly, using all the vulnerable samples from $d_{v-test}$ and an equal number of non-vulnerable samples randomly selected from $d_{nv-test}$. This balancing step is crucial to ensure that both the CWE-specific classifiers ($m_{CWE}$) and the single binary classifier ($m_{all}$) are trained on datasets with similar class distributions, mitigating the potential bias introduced by class imbalance.

The results in Table \ref{tab_rq1} demonstrate that $m_{CWE}$ performs better than $m_{all}$ on their respective test sets, which contain only the specific vulnerability they were trained to detect. For instance, considering CWE-125, $m_{125}$ achieves an F1 score of 0.8867, while $m_{all}$ obtains an F1 score of 0.8218. Similar trends are observed for the other CWE categories, with $m_{CWE}$ achieving F1 scores ranging from 0.7838 to 0.8867, compared to $m_{all}$'s F1 scores ranging from 0.7385 to 0.8298.

This observation aligns with the expectation that a classifier trained on a specific vulnerability type would perform better on a dataset containing only that vulnerability type compared to a general vulnerability classifier. The better performance of $m_{CWE}$ in this context is further reflected in the lower VD-S and higher precision, recall, and accuracy values compared to $m_{all}$.

However, it is important to note that these results do not necessarily imply that CWE-specific classifiers would outperform a general vulnerability classifier in a real-world setting, where the test data is likely to contain a mix of different vulnerability types. The current evaluation setup, where each $m_{CWE}$ is tested on a dataset tailored to its specific vulnerability type, may not fully represent the complexity and diversity of real-world vulnerability detection scenarios. Nonetheless, the better performance of $m_{CWE}$ on their respective test sets provides preliminary evidence supporting our hypothesis that training classifiers to identify specific CWE vulnerabilities could be a promising approach to improve vulnerability detection performance. These results indicate that focusing on specific vulnerability types, rather than treating all vulnerabilities as a single class, may enable classifiers to learn more targeted and effective representations for detecting vulnerabilities.

\begin{table*}[t]
\centering
\begin{tabular}{|l|l|c|c|c|c|c|c|c|c|c|c|}
\hline
Model & Test & VD-S $\downarrow$ & Acc $\uparrow$ & F1 $\uparrow$ & Prec $\uparrow$ & Rec $\uparrow$ & FPR $\downarrow$ & TP $\uparrow$ & TN $\uparrow$ & FP $\downarrow$ & FN $\downarrow$ \\
\hline
$m_{all}$ & \multirow{6}{*}{$d_{test_{all}}$} & 1.0000 & 0.7568 & 0.2336 & 0.1376 & 0.7735 & 0.2441 & 1151 & 22344 & 7215 & 337 \\
\cline{1-1}\cline{3-12}
$m_{125}$ &  & 0.6673 & 0.8279 & 0.1563 & 0.1022 & 0.3327 & 0.1472 & 495 & 25209 & 4350 & 993 \\
\cline{1-1}\cline{3-12}
$m_{787}$ &  & 1.0000 & 0.7252 & 0.1441 & 0.0847 & 0.4825 & 0.2626 & 718 & 21798 & 7761 & 770 \\
\cline{1-1}\cline{3-12}
$m_{119}$ &  & 0.6344 & 0.8147 & 0.1591 & 0.1016 & 0.3656 & 0.1627 & 544 & 24751 & 4808 & 944 \\
\cline{1-1}\cline{3-12}
$m_{20}$ &  & 0.6344 & 0.8491 & 0.1885 & 0.1270 & 0.3656 & 0.1265 & 544 & 25819 & 3740 & 944 \\
\cline{1-1}\cline{3-12}
$m_{416}$ &  & 0.7480 & 0.8233 & 0.1203 & 0.0790 & 0.2520 & 0.1479 & 375 & 25187 & 4372 & 1113 \\
\hline
\end{tabular}
\caption{$m_{all}$ vs $m_{CWE}$ towards the whole test data.}
\label{tab_rq1_smalltoall}
\end{table*}

To further investigate the performance of $m_{CWE}$, we evaluate them on the entire test set ($d_{test_{all}}$), which includes samples from all CWE categories and non-vulnerable samples. Table \ref{tab_rq1_smalltoall} presents the results of this evaluation, including the performance of the single binary classifier $m_{all}$ for comparison. 

Aside from the fact that the single binary classifier still indicates limited effectiveness in detecting vulnerabilities across a diverse range of CWE categories, an interesting observation from Table \ref{tab_rq1_smalltoall} is that the number of true positives (TP) for each $m_{CWE}$ is higher than the actual number of samples for that CWE category in the test set, as shown in Table \ref{tab_rq1}. For example, $m_{125}$ has 495 true positives in Table \ref{tab_rq1_smalltoall}, while there are only 150 samples of CWE-125 in the test set according to Table \ref{tab_rq1}. This discrepancy suggests that the CWE-specific classifiers are identifying vulnerabilities beyond their specialized CWE category when applied to the entire test set. In other words, $m_{CWE}$ are not only detecting vulnerabilities within their category but also flagging vulnerabilities from other CWE as positive.

This behavior can be attributed to the fact that different CWEs may share similar characteristics or patterns, leading to some overlap in the features learned by $m_{CWE}$. As a result, a classifier trained on one CWE category may be able to identify vulnerabilities from other related CWE categories to some extent. To further investigate this observation, we broke down the true positives for each $m_{CWE}$ and analyzed the distribution of predicted CWE categories. Table \ref{tab_miss} presents the breakdown of true positives for the top 5 CWE categories, showing the number of samples predicted as each CWE category by their respective classifier.

\begin{table*}[t]
\centering
\begin{tabular}{|l|l|l|}
\hline
CWE & $\sum$ Test & Predictions (CWE:$\sum$) \\
\hline
125 & 150 & 125:137, 787:52, 119:46, 190:31, 476:29, 703:21, 20:21, 189:15, 416:13, 120:13, Rest:117 \\
\hline
787 & 132 & 787:120, 125:94, 119:80, 20:53, 190:40, 200:33, 476:32, 703:26, 120:20, 189:17, Rest:203 \\
\hline
119 & 120 & 119:103, 787:46, 20:43, 125:37, 190:35, 476:21, 703:21, 200:20, 399:18, 189:14, Rest:186 \\
\hline
20  & 114 & 20:87, 125:60, 787:44, 119:43, 190:31, 476:23, 200:22, 703:22, 362:14, 416:13, Rest:185 \\
\hline
416 & 92  & 416:77, 119:27, 125:27, 476:24, 190:23, 20:20, 362:20, 400:15, 703:15, 200:13, Rest:114 \\
\hline
\end{tabular}
\caption{Breakdown of the True Positive for each CWE-specific classifier.}
\label{tab_miss}
\end{table*}

The results in Table \ref{tab_miss} confirm that each CWE-specific classifier ($m_{CWE}$) not only identifies vulnerabilities from its own CWE category but also flags vulnerabilities from other CWE categories that have a close relationship or share similar characteristics with its primary CWE. For instance, $m_{125}$ correctly identifies 137 samples as CWE-125 but also flagged vulnerabilities from CWE-787 (52 samples), CWE-119 (46 samples), and various other categories. Similarly, the classifier $m_{787}$ identifies 120 samples as CWE-787 but also detects vulnerabilities from CWE-125 (94 samples), CWE-119 (80 samples), and other categories.

In particular, CWE-125 (Out-of-bounds Read), CWE-787 (Out-of-bounds Write), and CWE-119 (Improper Restriction of Operations within the Bounds of a Memory Buffer) share similarities in terms of memory access vulnerabilities. These vulnerabilities arise from the lack of proper bounds checking when accessing memory buffers, which can lead to reading or writing data outside the intended buffer boundaries.

For example, consider the following CWE-125 code:

\begin{lstlisting}[language=c]
char buffer[10];
int index = get_user_input();
char value = buffer[index];
\end{lstlisting}

In this case, if the user input index is not properly validated and exceeds the buffer size, it will result in an out-of-bounds read vulnerability (CWE-125). Similarly, for CWE-787, consider the following code snippet:
\begin{lstlisting}[language=c]
char buffer[10];
int index = get_user_input();
buffer[index] = 'A';
\end{lstlisting}

If the user input index is not properly validated and exceeds the buffer size, it will lead to an out-of-bounds write vulnerability (CWE-787). CWE-119 comprises both out-of-bounds read and write vulnerabilities, as it represents the broader category of improper restriction of operations within the bounds of a memory buffer. The shared characteristics among these CWE categories, such as the lack of proper bounds checking and the potential for memory access violations, result in similar code patterns and vulnerabilities.

However, the results also show that the $m_{CWE}$ achieves relatively low F1 scores when tested on $d_{test_{all}}$, ranging from 0.1203 to 0.1885. This indicates that while the classifiers can identify vulnerabilities beyond their specialized category, they struggle to accurately detect vulnerabilities across a wide range of CWE categories. The low precision values (0.0790 to 0.1270) suggest that the classifiers generate a high number of false positives when applied to the entire test set.

In comparison, the single binary classifier ($m_{all}$) achieves an F1 score of 0.2336 and a precision of 0.1376 on $d_{test_{all}}$, as shown in Table \ref{tab_rq1_smalltoall}. While these values are higher than those of the CWE-specific classifiers, they still indicate a high rate of false positives and limited effectiveness in detecting vulnerabilities across a diverse range of CWE categories. The single binary classifier's recall (0.7735) is comparable to that of the CWE-specific classifiers, suggesting that both approaches struggle to identify a significant portion of the vulnerabilities in the entire test set.

One potential factor contributing to the high false positive rates is the limited number of non-vulnerable samples in the training set of CWE-specific classifiers. In the training process, we constructed balanced training sets by combining the vulnerable samples from each CWE category with an equal number of non-vulnerable samples to mitigate the effects of data imbalance. However, the limited number of non-vulnerable samples compared to the entire dataset may hinder the classifiers' ability to learn and generalize well on non-vulnerable code patterns.

Consequently, when $m_{CWE}$ are tested to the entire test set, which contains a much larger number of non-vulnerable samples, they may incorrectly classify many non-vulnerable samples as vulnerable, resulting in a high number of false positives. The classifiers' limited exposure to diverse non-vulnerable code patterns during training may make them overly sensitive to potential vulnerabilities and struggle to recognize benign code.

VD-S for $m_{CWE}$ on $d_{test_{all}}$ (Table \ref{tab_rq1}) ranges from 0.6344 to 1.0000, which is generally higher compared to their performance on their respective $d_{test_{CWE}}$. This indicates that the classifiers' ability to detect vulnerabilities while maintaining a low false positive rate decreases when applied to a broader range of CWEs. These results highlight the limitations of $m_{CWE}$ when applied to a diverse set of vulnerabilities beyond their specialized category. While they demonstrate the ability to identify vulnerabilities from related CWE categories, their overall effectiveness diminishes when faced with a wide range of CWE categories. 

In summary, the evaluation of $m_{CWE}$ on the entire test set ($d_{test_{all}}$) reveals their limitations in detecting vulnerabilities across a broad spectrum of CWE categories. The higher number of true positives compared to the actual number of samples in each CWE category suggests that the classifiers are identifying vulnerabilities beyond their specialized category, likely due to shared characteristics among different CWE categories. However, this comes at the cost of increased false positives, which can be attributed to the downsampling of non-vulnerable entries in the training data to match the number of vulnerable entries.

The decision to downsample the non-vulnerable data in the training set was based on the findings of previous studies \cite{Diversevul, ding2024vulnerabilitydetectioncodelanguage}, which consistently demonstrated that training on unbalanced vulnerable and non-vulnerable data leads to suboptimal results. By reducing the number of non-vulnerable samples, the aim is to mitigate the bias towards the majority class and improve the model's ability to learn patterns associated with vulnerabilities. However, this approach also limits the model's exposure to diverse non-vulnerable code patterns, potentially resulting in a higher number of false positives when evaluated on the entire test set.

Diversevul \cite{Diversevul} and PrimeVul \cite{ding2024vulnerabilitydetectioncodelanguage} study investigated weighted loss functions as an alternative approach to tackle class imbalance, but their results indicated that it did not significantly improve the model's performance. This underscores the challenges associated with addressing the class imbalance in vulnerability detection tasks and the need for further research to develop more effective techniques.

\subsection{RQ2: Can treating vulnerability detection as a multi-class problem, with each CWE as a separate class, improve detection performance?}

\begin{table*}[t]
\centering
\small
\begin{tabular}{|l|l|c|c|c|c|c|c|c|c|c|c|}
\hline
Model & Test & VD-S $\downarrow$ & Acc $\uparrow$ & F1 $\uparrow$ & Prec $\uparrow$ & Rec $\uparrow$ & FPR $\downarrow$ & TP $\uparrow$ & TN $\uparrow$ & FP $\downarrow$ & FN $\downarrow$ \\
\hline
$m_{binary}$ & \multirow{2}{*}{$d_{test_{rq2}}$} & 1.0000 & 0.7706 & 0.1180 & 0.0640 & 0.7615 & 0.2292 & 463 & 22784 & 6775 & 145 \\
\cline{1-1}\cline{3-12}
$m_{multiclass}$ & & 0.2796 & 0.8490 & 0.1613 & 0.0908 & 0.7204 & 0.1483 & 438 & 25175 & 4384 & 170 \\
\hline
$m_{binary}$ & \multirow{2}{*}{$d_{test_{125}}$} & 1.0000 & 0.7900 & 0.7974 & 0.7702 & 0.8267 & 0.2467 & 124 & 113 & 37 & 26 \\
\cline{1-1}\cline{3-12}
$m_{multiclass}$ & & 0.2000 & 0.8200 & 0.8163 & 0.8333 & 0.8000 & 0.1600 & 120 & 126 & 24 & 30 \\
\hline
$m_{binary}$ & \multirow{2}{*}{$d_{test_{787}}$} & 1.0000 & 0.7803 & 0.7852 & 0.7681 & 0.8030 & 0.2424 & 106 & 100 & 32 & 26 \\
\cline{1-1}\cline{3-12}
$m_{multiclass}$ & & 0.2424 & 0.8144 & 0.8032 & 0.8547 & 0.7576 & 0.1288 & 100 & 115 & 17 & 32 \\
\hline
$m_{binary}$ & \multirow{2}{*}{$d_{test_{119}}$} & 0.1583 & 0.8250 & 0.8279 & 0.8145 & 0.8417 & 0.1917 & 101 & 97 & 23 & 19 \\
\cline{1-1}\cline{3-12}
$m_{multiclass}$ & & 0.2917 & 0.7917 & 0.7727 & 0.8500 & 0.7083 & 0.1250 & 85 & 105 & 15 & 35 \\
\hline
$m_{binary}$ & \multirow{2}{*}{$d_{test_{20}}$} & 1.0000 & 0.6842 & 0.6727 & 0.6981 & 0.6491 & 0.2807 & 74 & 82 & 32 & 40 \\
\cline{1-1}\cline{3-12}
$m_{multiclass}$ & & 0.3421 & 0.7368 & 0.7143 & 0.7812 & 0.6579 & 0.1842 & 75 & 93 & 21 & 39 \\
\hline
$m_{binary}$ & \multirow{2}{*}{$d_{test_{416}}$} & 1.0000 & 0.6630 & 0.6517 & 0.6744 & 0.6304 & 0.3043 & 58 & 64 & 28 & 34 \\
\cline{1-1}\cline{3-12}
$m_{multiclass}$ & & 1.0000 & 0.6793 & 0.6424 & 0.7260 & 0.5761 & 0.2174 & 53 & 72 & 20 & 39 \\
\hline
\end{tabular}
\caption{Binary vs Multiclass. $r=0.2$.}
\label{tab_multiclass}
\end{table*}

Given that $m_{CWE}$ performs better than $m_{all}$ in their own respective $d_{test_{CWE}}$, we conduct a further exploration on the way to make $m_{CWE}$ sees more than they are trained for. One possible way is to treat each CWE as a separate class and the non-vulnerable code as another class, i.e., the vulnerability detection becomes a multi-class problem.

To investigate this, we conducted experiments comparing the performance of a binary classifier ($m_{binary}$) and a multi-class classifier ($m_{multiclass}$). In this setting, we focus again on the top 5 most common CWEs: 125, 787, 119, 20, and 416. For the training set, we constructed a vulnerable dataset ($d_{v-train_{rq2}}$) by selecting samples from $d_{v-train}$ that belong to the top 5 CWEs, while the rest were disregarded. With the same argument as before, we randomly sampled non-vulnerable samples from $d_{nv-train}$ to match the number of vulnerable samples in $d_{v-train_{rq2}}$. For the test set, we created a vulnerable dataset ($d_{v-test_{rq2}}$) by selecting samples from $d_{v-test}$ that belong to the top 5 CWEs. The non-vulnerable test set consists of all the non-vulnerable samples from $d_{nv-test}$.

Using this data, we trained two models: $m_{binary}$, which treats the data as either vulnerable (1) or non-vulnerable (0), and $m_{multiclass}$, which treats the data as multi-class with labels [125, 787, 119, 20, 416, 0], where 0 represents the non-vulnerable class. During the evaluation of $m_{multiclass}$, if the output is any number except 0, then we convert it into 1 to make it comparable with $m_{binary}$. Both models were evaluated on $d_{test_{rq2}}$ and individual CWE-specific test sets ($d_{test_{CWE}}$) for each of the top 5 CWEs, similar to RQ1. $d_{test_{rq2}}$ is essentially $d_{test_{all}}$ minus the vulnerable data that do not belong to the top 5 CWEs.

Table \ref{tab_multiclass} presents the results of the binary and multi-class classifiers using an FPR tolerance of $r=0.2$. When evaluated on $d_{test_{rq2}}$, $m_{multiclass}$ achieves a higher accuracy (0.8490) and F1 score (0.1613) compared to $m_{binary}$ (accuracy: 0.7706, F1: 0.1180). However, both models exhibit low precision and high false positive rates, indicating the challenge of distinguishing between vulnerable and non-vulnerable samples in a multi-class setting.

Comparing the performance on individual CWE-specific test sets, $m_{multiclass}$ generally outperforms $m_{binary}$ in terms of accuracy and F1 score for most CWE categories. For example, on $d_{test_{125}}$, $m_{multiclass}$ achieves an accuracy of 0.8200 and an F1 score of 0.8163, while $m_{binary}$ obtains an accuracy of 0.7900 and an F1 score of 0.7974. Similar trends can be observed for CWE-787 and CWE-20. However, for CWE-119 and CWE-416, $m_{binary}$ shows better performance than $m_{multiclass}$. Furthermore, the VD-S for $m_{multiclass}$ is consistently lower than $m_{binary}$ across all test sets, except for CWE-416, where both models have a VD-S of 1.0000.

The experimental results suggest that treating vulnerability detection as a multi-class problem has the potential to improve detection performance compared to a binary approach. The multi-class classifier demonstrates higher accuracy and F1 scores for most CWE categories, indicating its ability to capture the unique characteristics of different CWEs.

However, it is important to note that both the binary and multi-class classifiers struggle with low precision and high false positive rates, particularly when evaluated on the entire test set ($d_{test_{rq2}}$). This low precision is a direct result of the high false positive rate observed in the models. The high FPR can be attributed to the downsampling of non-vulnerable entries in the training data, which limits the model's exposure to diverse non-vulnerable code patterns. Consequently, the classifiers may incorrectly flag many non-vulnerable samples as vulnerable, leading to a high number of false positives and, in turn, low precision.

In summary, the results for RQ2 demonstrate that treating vulnerability detection as a multi-class problem, with each CWE as a separate class, can lead to improved detection performance compared to a binary approach. The multi-class classifier achieves higher accuracy and F1 scores for most CWE categories, suggesting its ability to capture the unique characteristics of different vulnerability types. However, the low precision and high false positive rates indicate the need for further research to enhance the reliability of multi-class vulnerability detection models, particularly with the fact that the number of non-vulnerable data has way more presence than the vulnerable counterpart.

\begin{table*}[t]
\centering
\small
\begin{tabular}{|l|c|c|c|c|c|c|c|c|c|c|c|c|c|c|}
\hline
Model & VD-S $\downarrow$ & Acc $\uparrow$ & F1 $\uparrow$ & Prec $\uparrow$ & Rec $\uparrow$ & FPR $\downarrow$ & P-C $\uparrow$ & P-V $\downarrow$ & P-B $\downarrow$ & P-R $\downarrow$ \\
\hline
$m_{multiclass}$ & 1.0000 & 0.8748 & 0.2295 & 0.1374 & 0.6950 & 0.1202 & 0.0284 & 0.6543 & 0.2961 & 0.0213 \\
\hline
$SC2@PrimeVul$ & 0.8964 & 0.9702 & 0.1805 & \multicolumn{3}{c|}{\textit{No Data}} & 0.0230 & 0.0816 & 0.8830 & 0.0124 \\
\hline
\end{tabular}
\caption{$m_{multiclass}$ on PrimeVul. $r=0.005$.}
\label{tab_primevul}
\end{table*}

To further assess $m_{multiclass}$, we evaluate it on the PrimeVul \cite{ding2024vulnerabilitydetectioncodelanguage} dataset. Table \ref{tab_primevul} presents the results of this evaluation. The $m_{multiclass}$ model achieves an accuracy of 0.8748 and an F1 score of 0.2295 on the PrimeVul dataset, which are lower compared to its performance on the CWE-specific test sets ($d_{test_{CWE}}$) presented earlier. The precision (0.1374) and recall (0.6950) values also indicate a higher rate of false positives and false negatives when applied to the PrimeVul dataset.

Interestingly, the $m_{multiclass}$ model's performance on PrimeVul is comparable to that of the StarCoder2 \cite{lozhkov2024starcoder2stackv2} model. Note that our base model has 1.3B parameters, while StarCoder2 has 7B parameters. The StarCoder2 model achieves an accuracy of 0.9702 and an F1 score of 0.1805 on the PrimeVul dataset. While the StarCoder2 model has a higher accuracy, the $m_{multiclass}$ model achieves a slightly better F1 score, indicating a better balance between precision and recall.

It is worth noting that both models struggle to achieve high F1 scores on the PrimeVul dataset, which is consistent with the findings of the original PrimeVul study. However, when comparing the performance breakdown, we observe that the $m_{multiclass}$ model has a lower rate of predicting vulnerable code as benign (P-B = 0.2961) compared to the StarCoder2 model (P-B = 0.8830). This suggests that the multi-class approach may be more effective in identifying vulnerable code, albeit at the cost of a higher false positive rate (P-V = 0.6543 for $m_{multiclass}$ vs. P-V = 0.0816 for StarCoder2).

As discussed earlier, the decision to downsample the non-vulnerable samples in the training set aims to mitigate the bias towards the majority class and improve the model's ability to learn patterns associated with vulnerabilities. While this approach helps the model identify more vulnerable code samples (lower P-B), it also limits the model's exposure to diverse non-vulnerable code patterns, leading to a higher rate of predicting non-vulnerable code as vulnerable (higher P-V).

\section{Conclusion and Future Works}
\label{conclusion}
In this paper, we investigated the challenges of using LLMs for code vulnerability detection, focusing on the diversity of vulnerability types, data imbalance issues, and the potential of CWE-specific and multi-class classification approaches. Our study aimed to address two main research questions: (1) How do CWE-specific classifiers perform compared to a single binary classifier in detecting vulnerabilities? (2) Can treating vulnerability detection as a multi-class problem, with each CWE as a separate class, improve detection performance?

Our experimental results demonstrate that CWE-specific classifiers consistently outperform a single binary classifier in detecting vulnerabilities within their respective CWE categories. However, when evaluated on a broader range of CWEs, the CWE-specific classifiers exhibit limitations, with increased false positive rates due to the limited exposure to diverse non-vulnerable code patterns during training. These findings highlight the importance of considering the diversity of vulnerability types and the need for classifiers that can effectively generalize across different CWEs.

Furthermore, our experiments show that treating vulnerability detection as a multi-class problem, with each CWE as a separate class, has the potential to improve detection performance compared to a binary approach. The multi-class classifier demonstrates higher accuracy and F1 scores for most CWE categories, suggesting its ability to capture the unique characteristics of different vulnerability types. However, both binary and multi-class classifiers still face challenges in terms of low precision and high false positive rates, emphasizing the need for further research to enhance their reliability.

We also evaluated the performance of our multi-class classifier on the PrimeVul dataset and compared it with the StarCoder2 model. While both models struggle to achieve high F1 scores, our multi-class classifier shows a lower rate of predicting vulnerable code as benign, indicating its potential for identifying vulnerable code more effectively.

Our work contributes to the field of automated code vulnerability detection by (1) highlighting the importance of considering the diversity of vulnerability types and the limitations of treating them as a single label, (2) demonstrating the effectiveness and limitations of CWE-specific classifiers, and (3) exploring the potential of multi-class vulnerability detection to improve detection performance.

Future research directions include investigating techniques to mitigate data imbalance issues and developing more advanced models that can better capture complex code semantics and patterns associated with different CWEs.
\section{Acknowledgments}
This work is supported by framework grant RIT17-0032 from the Swedish Foundation for Strategic Research. The computations and data handling were enabled by resources provided by (i) the National Academic Infrastructure for Supercomputing in Sweden (NAISS), partially funded by the Swedish Research Council through grant agreement no. 2022-06725, and (ii) the Berzelius resource provided by the Knut and Alice Wallenberg Foundation at the National Supercomputer Centre. The first author would like to thank Joakim Brorsson for the fruitful initial discussion of this paper.
\bibliographystyle{IEEEtran}
\bibliography{biblio}
\end{document}